\documentclass[a4paper,12pt,twoside]{article}
\usepackage{epsfig}
\usepackage{amssymb}
\usepackage{cite}

\usepackage{fancyheadings}
\advance \headheight by 3.0truept       
 
\newlength{\nseparation}
\newcommand{\newc}{\newcommand}
\newc{\met}{$\not\!\!E_T$}
\newc{\be}{\begin{equation}}
\newc{\ee}{\end{equation}}
\newc{\br}{\begin{eqnarray}}
\newc{\er}{\end{eqnarray}}
\newc{\ba}{\begin{array}}
\newc{\ea}{\end{array}}
\newc{\bi}{\begin{itemize}}
\newc{\ei}{\end{itemize}}
\newc{\bn}{\begin{enumerate}}
\newc{\en}{\end{enumerate}}
\newc{\bc}{\begin{center}}
\newc{\ec}{\end{center}}
\newc{\ul}{\underline}
\newc{\ol}{\overline}
\newc{\eps}{\epsilon}
\newc{\bs}{$B_s\rightarrow \mu^+\mu^-$~}
\newc{\cbs}{{\cal B}(B_s\rightarrow\mu^+\mu^-)}
\newc{\dam}{\delta a^{\rm SUSY}_{\mu}}
\newc{\lt}{\left}
\newc{\rt}{\right}
\newc{\no}{\nonumber}
\newc{\nn}{\nonumber\\}
\newc{\ds}{\displaystyle}
\newc{\e}{\epsilon}
\newc{\eq}[1]{(\ref{#1})}
\newc{\imag}{{\rm Im}\,}
\newc{\real}{{\rm Re}\,}
\newc{\mev}{\mbox{MeV}}
\newc{\gev}{\mbox{GeV}}
\newc{\ov}[1]{\overline{#1}}
\newc{\ie}{{\it i.e.}}
\newc{\ra}{\rightarrow}

\newc{\gsim}{\lower.7ex\hbox{$\;\stackrel{\textstyle>}{\sim}\;$}}
\newc{\lsim}{\lower.7ex\hbox{$\;\stackrel{\textstyle<}{\sim}\;$}}
\newc{\sgn}{\ensuremath{\mbox{sgn}\,}}

\begin{document}
\tolerance=100000
\thispagestyle{empty}
\setcounter{page}{1}
hep-ph/0207026\hfill
{\rm FERMILAB-Pub-02/129-T}\\
\mbox{~}\hfill {\rm DAMTP-2002-79}\\
\mbox{~}\hfill {\rm Cavendish-HEP-02/07}\\
\mbox{~}\hfill
{\rm  June 2002} \\

\vspace*{1cm}

\begin{center}
\boldmath
  {\LARGE \bf Trilepton Events and $B_s\ra\mu^+\mu^-$:\\[3mm]
        No-lose for mSUGRA at the Tevatron?}
\unboldmath
\vspace{1cm}

{{\Large A. Dedes$^a$}, {\Large H.~K. Dreiner$^a$},   
{\Large U. Nierste$^b$}, {\large and} {\Large P. Richardson$^c$}} \\[3mm]
{\it $^a$Physikalisches Institut der Universit\"at Bonn, 
 Nu\ss allee 12, 
 D-53115 Bonn, Germany
} \\[3mm]

{\it $^b$ Fermi National Accelerator Laboratory, Batavia, IL
60510-500, USA\footnote{Fermilab is operated by URA under DOE contract
No.~DE-AC02-76CH03000.}}  \\[3mm] 

{\it $^c$ Cavendish Laboratory and DAMTP, University of Cambridge, UK} \\[4mm]
\end{center}

\vspace*{1cm}

\begin{abstract}{\small\noindent
    We study the Tevatron search potential for minimal supergravity
    (mSUGRA) and find two observables which reveal complementary
    information on the mSUGRA parameter space: the ``gold-plated''
    decays of charginos/neutralinos to trilepton final states and the
    rare decay \bs. For the universal gaugino mass, $M_{1/2}$ below
    250 GeV and for the universal scalar fermion mass, $M_0$ outside
    of 200-370 GeV we face a ``no-lose'' situation for the Tevatron:
    If $\tan\beta \lsim 30$, the Tevatron has a chance to see the
    trilepton events but not $B_s\to \mu^+\mu^-$ during Run IIa,
    whereas for larger $\tan\beta$ the trilepton events become
    invisible (at least with an integrated luminosity of 2 fb$^{-1}$)
    while $B_s\to \mu^+\mu^-$ is enhanced to an observable level.  For
    this study we perform an updated analysis of the trilepton
    signature which includes the full set of the decay matrix elements
    and spin correlations.  This leads to a new, more promising
    search reach for the Tevatron.}

\end{abstract}

\vspace*{\fill}
\newpage
\setcounter{page}{2}

\section{Introduction and Motivation}
\label{sec:intro}


Even after imposing R-parity conservation by hand, the minimal
supersymmetric standard model (MSSM) has an embarrassingly large
number of parameters.  The construction of a model with only a few
parameters is desirable for {\ae}sthetic and practical reasons.
The minimal supergravity (mSUGRA) scenario \cite{mSUGRA,Soni:1983rm}
is currently the most elegant solution to this problem.  In mSUGRA,
local supersymmetry (supergravity) is spontaneously broken in the
hidden-sector: the minimum of the scalar potential violates
supersymmetry.  The superpartner of the graviton, the spin 3/2
gravitino, acquires a mass, via this super-Higgs effect
\cite{Cremmer}.  Supersymmetry breaking is communicated to the
observable sector via gravity interactions.  The quantum gravitational
corrections are in principle non-renormalizable, but taking the
``flat'' limit where the Planck scale goes to infinity ($M_P=(8\pi
G_N)^{-1/2} \ra \infty)$ we are left with an effective theory which is
renormalizable.  The most general form of the low energy {\it scalar
  potential} after supersymmetry breaking \cite{Soni:1983rm} contains
the following free parameters:\footnote{If R-parity is conserved there
  is only one bilinear parameter $b_{ij}=b$.}
\begin{enumerate}
\item mass terms for the scalar particles, $m_{ij}$, 
\item trilinear couplings, $h_{ijk}$,
\item the bilinear terms, $b_{ij}$.
\end{enumerate}
These terms break supersymmetry softly in the sense that no quadratic
divergences appear in the theory \cite{Girardello:1981wz}.  The number
of parameters is dramatically reduced if the superpotential splits
into a hidden-sector and a visible sector part \cite{Nilles:1983ge}.
Then these parameters are all universal
\begin{equation}
m_{ij}=M_0 \,\delta_{ij},\qquad h_{ijk}=A_0 \,Y_{ijk}, \qquad
b_{ij}=B_0\,\delta_{ij}.
\end{equation}  
$B_0$ and the magnitude of the superpotential Higgs mixing parameter
$\mu$ are fixed by electroweak symmetry breaking \cite{Ibanez:fr}.
Outside of the scalar sector, there is a further universal soft
supersymmetry breaking parameter: the common gaugino mass, $M_{1/2}$.
In the Higgs sector there is an additional parameter: $\tan\beta$, the
ratio of the vacuum expectation values of the two neutral, CP-even
Higgs fields. We thus have a total of 5 parameters in mSUGRA
\begin{equation}
M_0,\,A_0,
\,M_{1/2},\,\sgn\mu,\,\tan\beta.
\end{equation}

The parameters $M_0,\,A_0,\,M_{1/2},$ are expected to be universal at
the {\it Planck scale}, where supersymmetry breaking is mediated from
the hidden-sector to the observable sector.  In order to obtain
weak-scale predictions we must use the renormalization group equations
(RGEs) to evolve the parameters from $M_P$ down to $M_W$.  In
addition, there could be unification at the grand unified (GUT) scale
$M_X={\cal O} (10^ {16}\,GeV)$ in an unknown simple group such as
$SU(5)$ or $SO(10)$.  This would strongly modify the RGEs between the
Planck scale and the GUT scale.  In order to avoid this model
dependence, we consider $M_0,\,A_0,\,M_{1/2},$ to be universal at the
GUT scale.  We thus neglect any effects of the running of the RGEs
between $M_P$ and $M_X$.  In our numerical analysis, $\tan\beta$ and
$\sgn\mu$ are fixed at the weak scale.

The mSUGRA model with universal parameters is a simple, well motivated
model for the low-energy supersymmetric spectrum.  If an observable in
this model deviates substantially from the standard model (SM), then
we would expect to find a similar deviation in a more general
non-universal case, modulo accidental cancellations. Thus we expect to
be able to obtain more general results already in this simplified
model. However, correlations between observables in mSUGRA such as
\bs\ versus $(g-2)_\mu$ \cite{sakis} are typically lost in the more
general case due to the additional parameter freedom.

After more than 25 years there is still no experimental evidence for
supersymmetry (SUSY).  In principle we expect two kinds of signatures:
1) an indirect signature via a deviation from a SM prediction or 2) a
direct detection of a supersymmetric particle.  It is the purpose of
this letter to demonstrate the complementarity of these two approaches
in two specific examples.  For the direct search we consider the
trilepton signature, which was first proposed in the context of
supersymmetry in
\cite{Dicus:1983cb,Baer:1985at,Baer:1986dv,Baer:1986vf}, and has since
become a standard search mode for supersymmetry production. We wish to
investigate the correlation of this signature in mSUGRA with the rare
$B_s$ meson decay \bs. We expect these two signatures to be observable
at Run~II of the Tevatron.

In the following we present the actual status of leptonic $B$ meson
decays.  We then proceed and perform a brand new analysis of the
multilepton signatures in mSUGRA.  We confirm the results of Matchev
and Pierce \cite{Matchev:1999yn,Matchev:1999nb}.  We then go beyond
this work to include the full matrix elements for the supersymmetric
decays, which were not included in ISASUSY at the time.  This leads to
a more promising discovery reach for Run II of the Tevatron.

\section{The decay \boldmath{\bs}}
In the SM, the leptonic decays $B_{s,d}\ra \ell^+\ell^{-}$ are
dominated by Z-boson penguin and box diagrams involving top quark
exchange. The decay amplitude is helicity suppressed and thereby
proportional to the lepton mass.  The resulting small branching ratios
are well below the upper bounds set by past and present experiments;
the current situation is summarized in Table~\ref{tab}.  The
uncertainty in the SM predictions \cite{Buchalla} of Table~\ref{tab}
mainly stems from the decay constants $f_{B_d}=(200\pm 30)$ MeV and
$f_{B_s}=(230\pm 30)$ MeV~\cite{fbs}. While the element $|V_{ts}|=
0.040\pm 0.002\;$ of the Cabibbo-Kobayashi-Maskawa (CKM) matrix
entering the $B_s$ decays is well-known from CKM unitarity, the $B_s$
decays suffer from an additional uncertainty due to $|V_{td}|$.  From
a global fit to the unitarity triangle one finds $0.15 \leq
|V_{td}/V_{ts}| \leq 0.23$ \cite{ckmfit}.  We use the $\overline{\rm
  MS}$ b-quark mass $m_b(m_b)=4.25$ GeV and assume a fixed\footnote{
  In the mSUGRA analysis below, $m_t(m_t)$ is not fixed due to the
  variation of the strong coupling constant $\alpha_s$ with the SUSY
  spectrum under the assumption of gauge coupling unification.
  However, the effect of this variation is small compared to the
  hadronic uncertainty.}  $\overline{\rm MS}$ value for the top quark
mass, $m_t(m_t)=167$ GeV.  This corresponds to a pole mass of
$m_t=175$ GeV.  There is an additional small uncertainty of $\pm
0.3\times 10^{-9}$ in ${\cal B}($\bs$\!)$ when scanning the top mass
in its experimental range. Since we are interested in
order-of-magnitude effects from supersymmetry, we can safely neglect
the hadronic uncertainty and work with the central values of the decay
constants and CKM elements.

\begin{table}[t]\begin{center}
\begin{tabular}{|c|c|c|c|}\hline
Channel & Expt. & Bound (90\% CL) & SM prediction \cite{Buchalla}\\ \hline 
$B_s \to e^+ e^-$ & L3~\cite{L3}  
                  & $<5.4\times 10^{-5}$ 
                  & $(8.9 \pm 2.3 )\times 10^{-14}$ \\ \hline
$B_s \to \mu^+ \mu^-$ & CDF~\cite{CDFbmumu} 
                  & $<2.0\times 10^{-6}$ 
                  & $(3.8\pm1.0) \times 10^{-9}$ \\ \hline
$B_s \to \tau^+ \tau^-$  & LEP~\cite{Nardi} 
                 & $<0.05$  
                 & $(8.2\pm 2.1) \times 10^{-7}$ \\ \hline
$B_d \to e^+ e^-$ & CLEO~\cite{CLEO} 
                  & $<8.3\times 10^{-7}$ 
                  & $(2.4\pm 0.7 \pm 0.7) \times 10^{-15}$\\ \hline
$B_d \to \mu^+ \mu^-$ & CLEO~\cite{CLEO} 
                  & $<6.1\times 10^{-7}$ 
                  & $(1.0\pm 0.3\pm 0.3) \times 10^{-10}$  \\ \hline
$B_d \to \tau^+ \tau^-$  & LEP~\cite{Nardi} 
                         & $<0.015$ 
                         & $(2.1 \pm 0.6\pm 0.6)\times 10^{-8}$ \\ \hline
\end{tabular}\end{center}
\caption{{\it The experimental status and the SM predictions for the 
  branching ratios ${\cal B}(B_{s,d}\to \ell^+\ell^{-}$). The error in
  the $B_s$ branching ratios mainly originates from the uncertainty in
  $f_{B_s}$, the two errors in the $B_d$ branching ratios correspond
  to the uncertainties in $f_{B_d}$ and $|V_{td}|$.  }}\label{tab}
\end{table}

As ${\cal B} (B \to \ell^+ \ell^-) \propto m_{\ell}^2$, the branching
ratio is largest for $\ell=\tau$.  Yet $\tau$-lepton searches are very
difficult at hadron colliders. Therefore the best mode to search for
new physics is \bs.  Still the decay to $\tau^+\tau^-$ has a very weak
bound and we believe that an analysis of the existing LEP data could
result in a stronger bound.  Leptonic branching ratios of $B_d$ mesons
are smaller by a factor of $|V_{td} /V_{ts}|^2\lsim 0.05$ than the
leptonic decays of $B_s$ mesons.  From Table~\ref{tab} we see that
CLEO~\cite{CLEO} has provided strong bounds for the electron and muon
final states. Again, a dedicated search for the final state $\tau$'s
remains to be performed. The BaBar and Belle experiments could improve
on Table~\ref{tab} as well.

 From now on we restrict ourselves to the decay mode \bs.  In the SM,
the decay mode \bs is experimentally challenging due to its small
branching ratio.  However, it has a quite distinctive and unique
signature.  During Run I of the Tevatron the CDF experiment has set an
upper bound at 95\% CL. \cite{CDFbmumu}  
\begin{equation}
\cbs< 2.6 \times 10^{-6}\,,
\end{equation}
based on an analysis using very tight cuts to achieve an essentially
background-free framework. When extrapolated to Run IIa, which
corresponds to an integrated luminosity of 2 fb$^{-1}$, this method
yields a single event sensitivity of $\cbs =1.0 \cdot 10^{-8}$
\cite{fnalrep}.  A recent analysis~\cite{Tanaka} found that CDF can
discover \bs in Run~IIb with an integrated luminosity of 15 fb$^{-1}$,
if $\cbs > 1.2 \times 10^{-8}$.  Currently further sophisticated
studies are underway, and better cuts may well improve the explorable
range of $\cbs$. The SM range will certainly be accessible to the LHC
experiments~\cite{LHCb}.

The helicity suppression gives the leptonic $B$ decays a very special
r\^ole in the search for new physics, because they are sensitive to
new chirality-flipping interactions like extended Higgs sectors with
enhanced Yukawa couplings. This scenario occurs in the
two-Higgs-doublet model (2HDM) and the MSSM, if $\tan \beta$ is large.
In particular no other observable in $B$ physics is more sensitive to
effects from non-standard neutral Higgs bosons than ${\cal B} (B \to
\ell^+ \ell^-)$.  In the type-II 2HDM the extra contribution to the
decay amplitude grows like $\tan^2\beta$ and is roughly of the same
size as the SM amplitude \cite{Nierste,Huang}. In the MSSM, however,
the amplitude involves three powers of $\tan \beta$, so that 
\begin{equation}
\cbs\propto \tan^6\beta. 
\end{equation}
This leads to an enhancement over the SM value by up to three orders
of magnitude.  This holds even in the flavour-universal MSSM, where
flavour changes occur in weak vertices only and come with the usual
CKM elements \cite{Babu,Huang,Urban}.
Moreover, in the limit of infinite SUSY mass parameters the MSSM
contribution does not vanish, because the MSSM approaches a general
2HDM with flavour-changing neutral Higgs couplings in this limit. This
feature and the resulting importance of $B$ physics to probe the large
$\tan\beta$ regime of the MSSM was first emphasized in
\cite{Maxim}. The supersymmetric contributions enhancing $\cbs$ will
also affect the branching ratio and forward-backward asymmetry in the
decay
\begin{equation}
B\ra (\pi,K) \ell^+\ell^-,
\end{equation}
but these effects are small and strongly correlated with $\cbs$ and
will only be measurable, if $\cbs$ is close to its experimental upper
bound \cite{Urban,Olive}.

The first analysis of \bs in the minimal supergravity scenario
(mSUGRA) revealed an interesting correlation of $\cbs$ with the muon
anomalous magnetic moment $(g-2)_\mu$ and the mass of the lightest
Higgs boson \cite{sakis}: If $(g-2)_\mu$ exceeds the SM prediction by
$40\times 10^{-10}$ (which is the maximum allowed from
BNL \cite{BNL,Narison}), $\;\cbs$ is larger by a factor of 10--100 than
in the SM and within reach of Run II of the Tevatron.  If the decay
\bs is observed at Run II of the Tevatron, then in mSUGRA the mass of
the lightest neutral CP-even Higgs boson should be less than 120 GeV.
The results of Ref.~\cite{sakis} were recently confirmed in
Ref.~\cite{Tanaka}, where also constraints from dark matter were
included.  A nice observation was made in Ref.~\cite{Baek}: alternative
SUSY breaking scenarios imprint signatures on $(g-2)_\mu$ and $\cbs$
which are very different from those found for mSUGRA in
Ref.~\cite{sakis}.  A substantial enhancement of $\cbs$ in
anomaly-mediated SUSY breaking scenarios is impossible and also hard
to accommodate with gauge-mediated SUSY-breaking.  Thus an observation
of \bs\ at Run II of the Tevatron can be viewed as a signature of
supersymmetry with gravity-mediated supersymmetry breaking. In our
numerical analysis of $\cbs$ we use the formulae of \cite{Urban}
supplemented with an all-order resummation of $\tan\beta$-enhanced 
supersymmetric QCD corrections \cite{cgnw}. Other QCD corrections
(neglecting gluino contributions) were calculated in \cite{Buras}
and found to be small, if the renormalization scale is chosen close to
the masses of the SUSY particles propagating in the loop. 


\section{Multilepton SUSY events}

Trileptons are an old phenomenological signature of cascade decays in
particle physics \cite{Barger:1977ap,Barger:1984rp}.  They were first
applied in supersymmetry in the cascade decays of electroweak gauge
bosons via light gauginos \cite{Dicus:1983cb}.\footnote{Implicitly
  these signatures appear in Ref.~\cite{Chamseddine:eg}.}  A first
systematic study was performed in
\cite{Baer:1985at,Baer:1986dv,Baer:1986vf}.  It was first shown in
\cite{Nath:sw} that off-shell $W,Z$ cascade decays via gauginos can
also lead to promising trilepton signatures.  The study was however
restricted to $\tan\beta=1$.  Also the sparticle spectrum was not
computed in terms of a small set of MSSM parameters.  These points
were generalized in \cite{Baer:1992dc}.  An extension of the Tevatron
studies to the LHC was performed in \cite{Baer:nr}\footnote{For
  analogous signatures for gluino production see for example
  \cite{Dreiner:ba}.} and to a $\sqrt{s}=4\,$TeV Tevatron in
\cite{Kamon:1994yq}. A first study for the high-luminosity Run II at
the Tevatron was performed in \cite{Baer:1995bu,Mrenna:1995ax}.  A
first study of what we now call mSUGRA was performed in
\cite{Baer:1994nc} including a multichannel analysis.  Various
unification scenarios were probed via the trilepton signature in
\cite{Baer:1994dq,Lopez:1994dm}.  In \cite{Baer:1997yi,Baer:1998bj}
the case of large $\tan\beta$ was first considered.  This can lead to
additional contribution of $\tau$ leptons in the final state which was
discussed in detail in
\cite{Barger:1998wn,Barger:1998hp,Baer:1998sz,Lykken:1999kp,Guchait:2002xh}.
An update improving many features of the analysis has recently been
performed in \cite{Matchev:1999yn,Baer:1999bq} including a
reevaluation of the $W\gamma^*$ background \cite{Matchev:1999nb}.  We
now discuss in detail how we have improved on this work.

\subsection{Numerical Analysis}
In general our strategy for the analysis of the multilepton signatures
closely follows that of Matchev and Pierce
\cite{Matchev:1999yn,Matchev:1999nb}. The major differences between
these studies are:
\begin{itemize}
\item While Matchev and Pierce \cite{Matchev:1999yn,Matchev:1999nb}
  used leading-order (in QCD) cross sections for both the signal and
  background processes we use next-to-leading order
  calculations for the background and most important signal processes.
    
\item In Refs.~\cite{Matchev:1999yn,Matchev:1999nb} the authors used
  ISAJET 7.42 \cite{Baer:1999sp}
  for the simulation of the signal. For the background
  they used PYTHIA 6.115 \cite{pythia}. In Ref.\cite{Matchev:1999nb}
  they used a combination of PYTHIA and COMHEP \cite{Pukhov:1999gg}
  for the WZ background.\footnote{This brief summary is based on a
    private communication with Konstantin Matchev and differs slightly
    from the presentation in the papers.}  Instead, we use HERWIG6.4
  \cite{HERWIG64} to simulate the signal and a combination of HERWIG
  and COMPHEP to simulate the background.  This has a number of
  advantages due to recent improvements in HERWIG. While ISAJET7.42 uses
  matrix elements for the three body decays of the gauginos
  it performs the production and decay of the SUSY particles
  independently. HERWIG6.4 makes use of the
  method described in \cite{Richardson:2001df} to perform the SUSY
  decays including all the momentum and spin correlations. One of the
  major advantages of our studies is that the spin correlation
  algorithm used in HERWIG can use the TAUOLA decay package
  \cite{Jadach:1993hs} to give the correct helicity of the tau on an
  event-by-event basis, rather than averaging as was done in
  \cite{Matchev:1999yn,Matchev:1999nb}. ISAJET7.58 is  used to 
  calculate the SUSY spectrum and decay rates.
\end{itemize}


\noindent In our simulation we make use of the PGS \cite{pgs} detector
simulation.  This is a more recent version of the SHW package.  The
default PGS parameters are used together with the following additions.
\begin{enumerate}
\item A requirement that the total transverse energy in a cone
      of size $\Delta R=\sqrt{\Delta\phi^2+\Delta\eta^2}=0.4$ about the
      direction of a muon is less than 2~GeV. Here $\phi$ is the
      azimuthal angle of the particle, $\eta=-\ln\tan(\theta/2)$ is the
      pseudo-rapidity, and $\theta$ is the angle of the particle with respect to 
      the incoming beam.
\item The missing transverse energy, \met, which is calculated in PGS using the
      calorimeter cells, is corrected to include any muons. The transverse energy
      $E_T=E\sin\theta$, where $E$ is the energy of a particle.
\item A cut on the ratio of the electromagnetic, $E_{\rm em}$,
      to hadronic energy, $E_{\rm had}$, in a jet $ E_{\rm em}/E_{\rm had}<10$
      to avoid electrons being considered as jets.
\end{enumerate}
These modifications are virtually identical\footnote{In
  Ref.~\cite{Matchev:1999nb} the tracking coverage is extended from
  $|\eta|\leq1.5$ to $|\eta|\leq2.0$ however this is now the default
  in PGS.} to those made in Ref.~\cite{Matchev:1999nb}.

\subsection{Background and Signal}
The major backgrounds to multilepton production are:
\begin{description}
\item[\underline{WW}:] Here the background is from the production of
  leptons in the cascade decay of the W bosons. These events are
  simulated using HERWIG6.4 and the results normalized to the
  next-to-leading order result from MCFM \cite{Campbell:1999ah}.
  
\item[\underline{WZ}:] In this case the leptons come from the decay of
  the gauge bosons.  This is the main physics background in most of
  the channels. As this channel is so important, rather than use
  HERWIG, which does not include all the diagrams, we use COMPHEP to
  generate all the Feynman diagrams for the four-fermion final state.
  HERWIG is then used to generate the initial-state radiation and
  hadronization of the partonic events generated with COMPHEP. The
  results are normalized to the next-to-leading order cross section
  calculated using MCFM \cite{Campbell:1999ah}.
  
\item[\underline{ZZ}:] The background comes from the production of
  leptons in the decay of the Z bosons. These events are simulated
  using HERWIG6.4 with a next-to-leading order normalization from MCFM
  \cite{Campbell:1999ah}.
    
\item[\underline{W+jets}:] In these events the background comes from
  one real lepton produced in the decay of the W and additional fake
  leptons from the mis-identification of isolated hadrons as leptons.
  The simulation of isolated hadrons producing fake electrons and
  muons cannot be performed reliably using Monte Carlo event
  generators and therefore we adopt the same procedure as in
  Ref.~\cite{Matchev:1999nb} in order to estimate this rate from the
  Run~I CDF data.  The SHW simulation, which was the predecessor to
  PGS, gives a reasonable simulation of fake tau production
  \cite{pgstau,Matchev:1999nb}, and we therefore rely on the Monte
  Carlo simulation for fake tau production.  HERWIG6.4, which includes
  the matrix-element correction to the parton shower as described in
  Ref.~\cite{Corcella:1999gs}, is used to simulate this background.
  The results of the Monte Carlo simulation are re-scaled to the
  next-to-leading-order cross section calculated using MCFM
  \cite{Campbell:2002tg}.
    
  \item[\underline{Z+jets}:] In this case the background comes from
    either the leptons produced in the cascade decay of the Z, with
    additional fake leptons as described above, charge
    mis-identification, or leptons from the Z itself depending on the
    channel.  Again HERWIG6.4 which includes the matrix-element
    correction to the initial-state radiation is used to simulate
    this process.  As with the W+jets background, the results of the
    simulation are normalized using the next-to-leading-order cross
    section calculated with MCFM \cite{Campbell:2002tg}. The fake
    lepton rate is calculated as for the W+jets events.
    
  \item[\underline{\boldmath{${\rm t}{\rm \bar{t}}$}}:] The background
    comes from the semi-leptonic decay of the top quarks. We simulate
    this background using HERWIG6.4 and use the next-to-leading-order
    with next-to-leading-log resummation calculation of
    Ref.~\cite{Bonciani:1998vc} to normalize the results.

\end{description}

\noindent There is also a potential background from the production of ${\rm
  b}{\rm \bar{b}}$ followed by the semi-leptonic decay of the produced
b-flavoured hadrons. However, this should be negligible after the
isolation and cuts on the transverse momentum, $p_T$, of the leptons.
  
In all cases we use HERWIG6.4 to simulate the signal.  The cross
sections for $\tilde{\chi}^0_2\tilde{\chi}^\pm_1$ and $\tilde{
  \chi}^+_1\tilde{\chi}^-_1$, which give the dominant contribution to
the SUSY production, are normalized to the next-to-leading order cross
section using the results of \cite{Beenakker:1999xh}.  The
leading-order cross section from HERWIG is used for the remaining
processes.

\subsection{Cuts}

The basic idea of the analysis is to take several different cuts,
which should differentiate between the signal and the background,
consider all possible combinations of the cuts at each point in
parameter space and optimize the choice of cuts in order to maximize
$S/\sqrt{B}$, where $S$ and $B$ are the number of signal and background
events respectively. In addition to the requirement of $S/\sqrt{B}>5$ all the
contours shown also require that there at least five signal events.

The cuts can be split into two types: channel-independent cuts and
those which are specific to a given channel.

The channel independent cuts we use are:
\begin{enumerate}
  
\item Five possible cuts on the missing transverse energy,
  $\not\!\!E_T> \{15,20,25,30\}$~GeV or no cut;
  
\item An optional veto on the presence of QCD jets with
  $E_T>\{10,15,20,25,30\}$~GeV in the event.
  
\item A cut on the minimum invariant mass of any opposite sign same
  flavour electron or muon (OSSF) pairs in the event
  $m_{\ell\ell}>\{5,10,15,20,25,30,35,40,45,50,55,60\}$~GeV, or no cut.
  
\item A cut on the maximum invariant mass of any OSSF pairs in the
  event \mbox{$m_{\ell\ell}<\{50,60,70,80\}$~GeV}, or a cut on the minimum
  difference between the mass of any OSSF pair and the Z mass $|M_{\rm
    Z}-m_{\ell\ell}|>\{10,15\}$~GeV, or no cut.
  
\item A cut removing any events which have an electron or muon with
  transverse mass $65\,\rm{GeV} \leq m_T \leq 85\,\rm{GeV}$. The
  transverse mass is given by
  \mbox{$m^2_T=2|p_{T\ell}||p_{T\rm{miss}}|(1-\cos\Delta\phi)$}, where
  $p_{T\ell}$ is the transverse momentum of the lepton,
  $p_{T\rm{miss}}$ is the missing transverse momentum and $\Delta\phi$
  is the azimuthal angle between the lepton and the missing transverse
  momentum.

\end{enumerate}

The channel dependent cuts are:
\begin{enumerate}
  
\item For the trilepton and like-sign lepton signature we require a
  central electron or muon with $|\eta|\leq1$ and $p_T>{11, 15, 20}$~GeV,
  or no requirement for a central lepton.
  
\item For the lepton+ tau jet channel we require that the tau decay
  only has one prong\footnote{\ie\ the tau jet only had one charged
    track.}  and that the fraction of the tau jet's momentum carried
  by the charged track is greater than 80\%, as suggested in
  \cite{Guchait:2001gk}.
  
\item A cut on the transverse momenta of the leptons is also applied
  depending on the channel, the cuts for the various different
  channels are given in Table~\ref{tab:ptcut}.

\end{enumerate}
These cuts are the same as used in
Ref.~\cite{Matchev:1999yn,Matchev:1999nb} {\it apart} from one extra missing
transverse energy cut and the cut on the momentum of the track in the
tau jets.

\begin{table}[tbh]
\begin{center}
\begin{tabular}{|c|c|c|c|}
\hline
\multicolumn{4}{|c|}{Trilepton channel}\\
\hline
Cut Set        & $p_T(\ell_1)$ & $p_T(\ell_2)$ & $p_T(\ell_3)$ \\
\hline
1       & 11 & 5  & 5 \\
2       & 11 & 7  & 5 \\
3       & 11 & 7  & 7 \\
4       & 11 & 11 & 11\\
5       & 20 & 15 & 10\\
\hline
\multicolumn{4}{|c|}{Like-sign dilepton channel}\\
\hline
 Cut Set        & $p_T(\ell_1)$ & $p_T(\ell_2)$&  \\
\hline
1      & 11 &  9 &\\
2      & 11 & 11 &\\
3      & 13 & 13 & \\
4      & 15 & 15 &\\
5      & 20 & 20 & \\
\hline
\multicolumn{4}{|c|}{Dilepton+hadronic tau channel}\\
\hline
Cut Set  & $p_T(\ell_1)$ & $p_T(\ell_2)$ & $p_T(\tau)$ \\
\hline
1     &  8  &  5  &10\\
2     &  8  &  5  &15\\
3     &  11 &  5  &10\\
4     &  11 &  5  &15\\
5     &  15 & 10  &15\\ 
\hline
\end{tabular}
\end{center}
\caption{{\it Lepton $p_T$ cuts, all the cuts are given in GeV.}}
\label{tab:ptcut}
\end{table}

\begin{figure}[t]
\centerline{\hbox{\psfig{figure=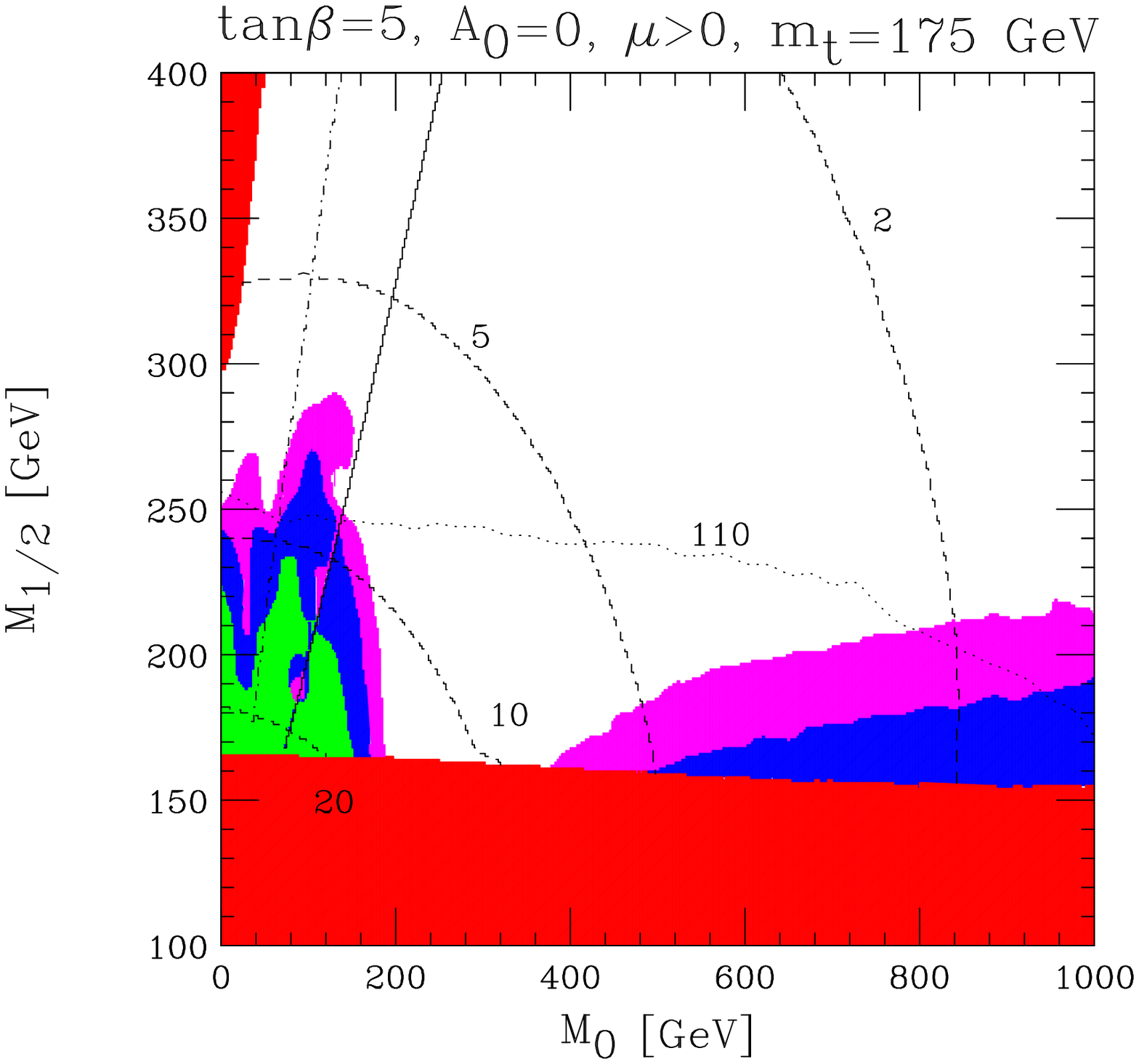,height=3in,angle=90}}
{\psfig{figure=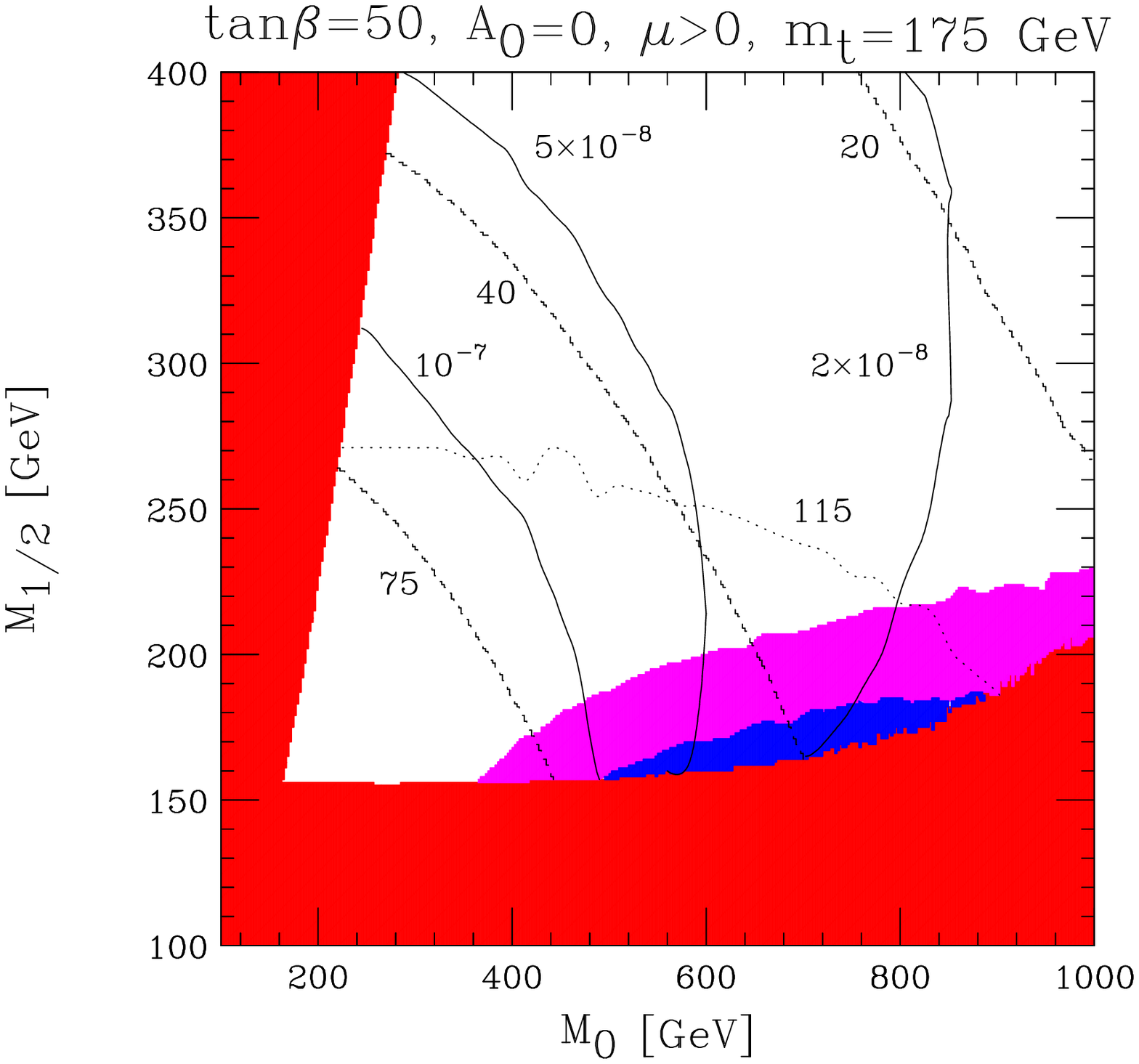,height=3in,angle=90}}}
\caption{{\it Tevatron 5$\sigma$ reach in the trilepton (3L) channel
in the $M_0-M_{1/2}$ plane in the mSUGRA scenario for small
$\tan\beta=5$ (left) and large $\tan\beta=50$ (right) and $A_0=0$,
$\mu>0$ and $m_t=175$ GeV. The shaded areas indicate the $5\sigma$
reach for an integrated luminosity of 30$fb^{-1}$ (magenta),
10$fb^{-1}$ (blue) and 2$fb^{-1}$ (green) and from top to bottom,
respectively. Dashed contours represent the SUSY contribution to the
muon anomalous magnetic moment (in units of $10^{-10}$) and the dotted
contours are iso-mass contours of the lightest neutral Higgs
boson. The solid contour (only for $\tan\beta=50$) indicates the
prediction for the branching ratio $\cbs$.  In the left
($\tan\beta=5$) plot the solid line indicates where
$m_{\tilde{\chi}^\pm_1}=m_{\tilde{\tau}_1}$ and the dot-dashed line
$m_{\tilde{\chi}^\pm_1}=m_{\tilde{\nu}_\tau}$.  The large red regions
are excluded by theory and experiment \cite{sakis}.}}
\end{figure}

\section{Discussion}
\subsection{Trilepton (3L) Channel}

We first discuss the trilepton (3L) channel.  The production cross
section for the chargino-neutralino pair, $\tilde{\chi}_1^\pm
\tilde{\chi_2}^0$, scales like $M_{1/2}^{-11/2}$ and is thus only
relevant in regions with small $M_{1/2}$, {\it i.e.} $M_{1/2}\lsim
300$ GeV. We present our results in Fig.~1 on the $M_0-M_{1/2}$ plane,
for two different values of $\tan\beta$: 5 and 50. Starting with the
(3L) events in the small $\tan\beta$ plot, we observe that the
Tevatron could reach a 5$\sigma$ discovery depending on its luminosity
in the regions
\begin{eqnarray}
{\rm Region~I}~(3L, \tan\beta=5) &:& \Biggl \{\begin{array}{c} 
M_0 \lsim 175~{\rm GeV} \;\;\;,\;\;\; M_{1/2} \lsim 230~{\rm GeV} 
\;\;\;,\;\;\; {\cal L}=2 fb^{-1} \;, \\
M_0 \lsim 190~{\rm GeV} \;\;\;,\;\;\; M_{1/2} \lsim 270~{\rm GeV} 
\;\;\;,\;\;\; {\cal L}=10 fb^{-1} \;, \\
M_0 \lsim 200~{\rm GeV} \;\;\;,\;\;\; M_{1/2} \lsim 290~{\rm GeV} 
\;\;\;,\;\;\; {\cal L}=30 fb^{-1} \;, \end{array} \nonumber \\[3mm]
{\rm Region~II}~(3L, \tan\beta=5) &:&\Biggl \{\begin{array}{c} 
M_0 \gsim 480~{\rm GeV} \;\;\;,\;\;\; M_{1/2} \lsim 190~{\rm GeV} 
\;\;\;,\;\;\; {\cal L}=10 fb^{-1} \;, \\
M_0 \gsim 360~{\rm GeV} \;\;\;,\;\;\; M_{1/2} \lsim 220~{\rm GeV} 
\;\;\;,\;\;\; {\cal L}=30 fb^{-1} \;, \end{array} \nonumber
\end{eqnarray}
Region~I is larger than in Ref.~\cite{Matchev:1999yn,Matchev:1999nb}
due to the extra missing transverse energy cut we have made.
Region~II is larger than in \cite{Matchev:1999yn,Matchev:1999nb} due
to the lighter spectrum produced by ISAJET7.58 which gives
a larger signal cross section than for the same SUGRA 
point with ISAJET7.42.  Overall,
we thus obtain a more promising trilepton search reach at Run~II of
the Tevatron than in previous studies.  This is one of the main
conclusions of this paper. 

Although we present our analysis in terms of the fundamental
parameters of mSUGRA $M_0,M_{1/2},\tan\beta$, there exist mass sum
rules~\cite{Martin} relating these parameters to the physical masses.
We can thus approximately translate the mSUGRA parameters into the
physical masses (up to a 10\% accuracy),
\begin{eqnarray}
2 m_{\tilde{\chi}_1^0} & \simeq & m_{\tilde{\chi}_2^0} 
\simeq m_{\tilde{\chi}_1^\pm} \simeq \frac{1}{3} m_{\tilde{g}} \simeq
0.8 M_{1/2} \;, \nonumber \\
m_{\tilde{q}} & \simeq & \sqrt{M_0^2+4 M_{1/2}^2} \;,
\end{eqnarray}
where $m_{\tilde{\chi}_i^0}$ are the neutralino masses with the
lightest being the LSP. $ m_{\tilde{\chi}_i^\pm}$ denote the chargino
masses.  $m_{\tilde{g}}$ and $m_{\tilde{q}}$ denote the gluino and the
heavy first and second generation squark masses, respectively. Thus
the Tevatron trilepton reach is relevant for a lightest chargino and a
next-to-lightest neutralino mass below 200 GeV and (indirectly) for a
gluino mass below 600 GeV.

In Fig.~1 we include Higgs boson mass isocurvatures.  The calculation
of the light Higgs boson mass strictly follows the procedure described
in Ref.\cite{FeynHiggs}.  In Region~I the light Higgs boson mass
varies from $(106 \div 112)\pm 3$ GeV where the error denotes a
theoretical error in the calculation~\cite{Sven}.  The current status
for the Higgs boson mass bound from LEP is~\cite{LEPHiggs}~: $M_h
\gsim 113.5~~{\rm GeV}$ for $\sin^2(\beta-\alpha)\simeq 1$, where
$\alpha$ is the mixing angle in the CP-even Higgs sector.  Thus
$\tan\beta=5$ has to be considered as the minimum $\tan\beta$ value
where trilepton events can appear at the Tevatron since lower values
will be in conflict with the LEP Higgs bound. The Higgs mass increases
with $\tan\beta$.

For small values of $\tan\beta$ the supersymmetric contribution to the
muon anomalous magnetic moment ($\dam$) is in general
small~\cite{Moroi}.  However, observing a trilepton signal in Region I
will imply an at least moderate enhancement of $a_\mu$ by 
\begin{equation}
\dam \gsim 7\times 10^{-10}\;,
\end{equation}
due to the light charginos and sneutrinos.  This is not the case in
Region~II, where $\,\dam$ approaches zero.  

There is a ``null'' $M_0-M_{1/2}$ region at the Tevatron where no (3L) signal
can be observed.  This ``null'' region spans the parameter space where
$200\lsim M_0 \lsim 370$ GeV and is due to the fact that off-shell,
slepton mediated decays of the gauginos destructively interfere with
the gauge boson mediated decays.  Since the decay $B_s\ra\mu^+\mu^-$
cannot be seen for such small values of $\tan\beta$ the ``null'' area
has to be covered by other Tevatron searches 
(possibly Higgs searches~\cite{Carena}).
  
In Region~II, a 5$\sigma$ signal reach can be achieved with ${\cal
  L}=10,30\, \rm{fb}^{-1}$.  The light Higgs boson mass varies between
$(106\div 111)\pm 3$ GeV and the supersymmetric contributions to the
muon anomalous magnetic moment are small,
\begin{equation}
\dam \lsim 5\times 10^{-10}.
\end{equation}

The CLEO/Belle/ALEPH allowed band for the branching ratio ${\cal
  B}(b\to s \gamma)$ does not set any constraint in the low
$\tan\beta=5$ region.  In what follows, we adopt the conservative
value\footnote{We follow here the discussion in the footnote no.~18 of
  Ref.\cite{Drees}.}  $2\times 10^{-4}\lsim {\cal B}(b\to s
\gamma)\lsim 5\times 10^{-4}$.

We now turn to the discussion of the large $\tan\beta=50$ plot in
Fig.~1.  
The new Region~I (low $M_0$)  is not there any more (see below).
The new Region~II is given by 
\begin{eqnarray}
{\rm Region~II}~(3L, \tan\beta=50) &:&\Biggl \{\begin{array}{c} 
M_0 \gsim 480~{\rm GeV} \;\;\;,\;\;\; M_{1/2} \lsim 185~{\rm GeV} 
\;\;\;,\;\;\; {\cal L}=10 fb^{-1} \;, \\
M_0 \gsim 360~{\rm GeV} \;\;\;,\;\;\; M_{1/2} \lsim 230~{\rm GeV} 
\;\;\;,\;\;\; {\cal L}=30 fb^{-1} \;. \end{array} \nonumber
\end{eqnarray}
The trilepton search only covers a
small range of the $M_0-M_{1/2}$ plane.  There is a further strong
constraint.  The light Higgs boson mass varies in the range $(110\div
119)\pm 3$ GeV from bottom to top where the contour $m_h=115$ GeV is
indicated in Fig.~1 as a dotted line.  The LEP preliminary bound, $m_h
\gsim 113$ GeV, is evaded (almost) everywhere in this plot.  However,
this is not true for the $b\to s \gamma$ constraint.  Using the
results in Ref.\cite{Drees} we find that the excluded region due to
the $b\to s \gamma$ constraint (accidentally) co\"{\i}ncides almost
exactly with the area below the $m_h=115$ GeV Higgs mass contour.  We
thus do not redraw it.  This leaves only a small area for the 3L
events to appear at the Tevatron
\begin{eqnarray}
{\rm Region~II'}~(3L, \tan\beta=50) &:& 
M_0 \gsim 800~{\rm GeV}\,; \;\;M_{1/2} \sim 190\div230~{\rm GeV},
\;\; {\cal L}=30 fb^{-1} \;.  \nonumber
\end{eqnarray}
Notice also that this area is bounded for large $M_0$ due to the
radiative electroweak symmetry breaking requirement.

By contrast, the decay mode \bs is very promising for $\tan\beta=50$.
The branching ratio varies from a handful of events for large $M_0$ up
to 100 events for low $M_0$ already with ${\cal L}=2 fb^{-1}$.  This
very nicely complements the trilepton search and is the second main
result of our paper: the tri-lepton search covers most of the low
$\tan\beta$ region, the branching ratio $\cbs$ covers most of the high
$\tan\beta$ region.

Fig.1 also verifies what had been seen in Ref.\cite{sakis}: the
branching $\cbs$ and the anomalous magnetic moment of the muon,
$\dam$, go hand-in-hand, {\it i.e.} there is a strong correlation.
New experimental results for the anomalous magnetic moment of the muon
are expected to appear as this manuscript is written.  The up-to-date
BNL excess is $-4.2\times 10^{-10} \lsim \dam \lsim 41.3 \times
10^{-10}$, at 90\% C.L ~\cite{BNL,Narison} with the central value
$\dam = 18 \times 10^{-10}$~\cite{Narison}.

\begin{figure}[t]
\centerline{\hbox{\psfig{figure=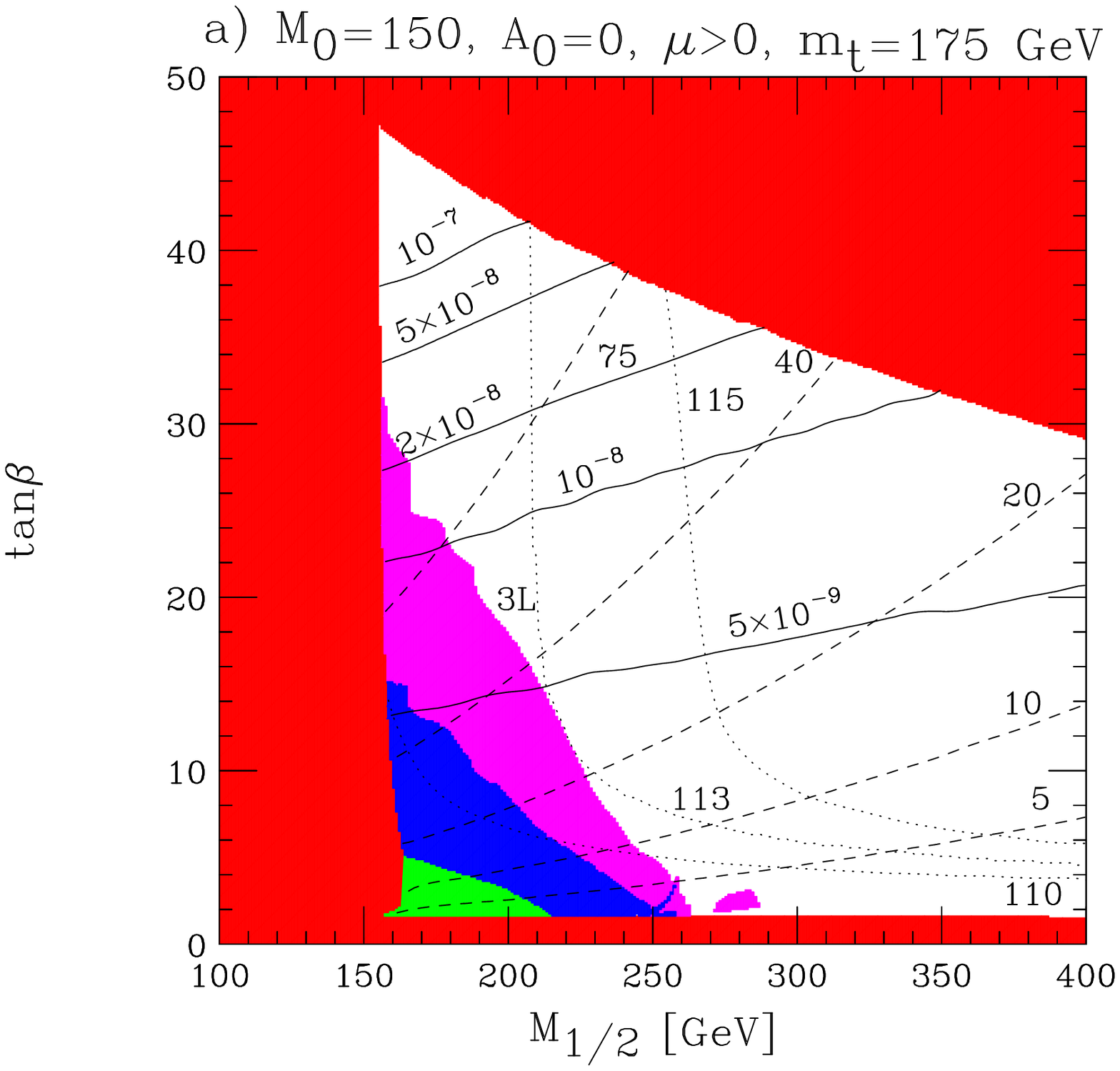,height=3in,angle=90}}
{\psfig{figure=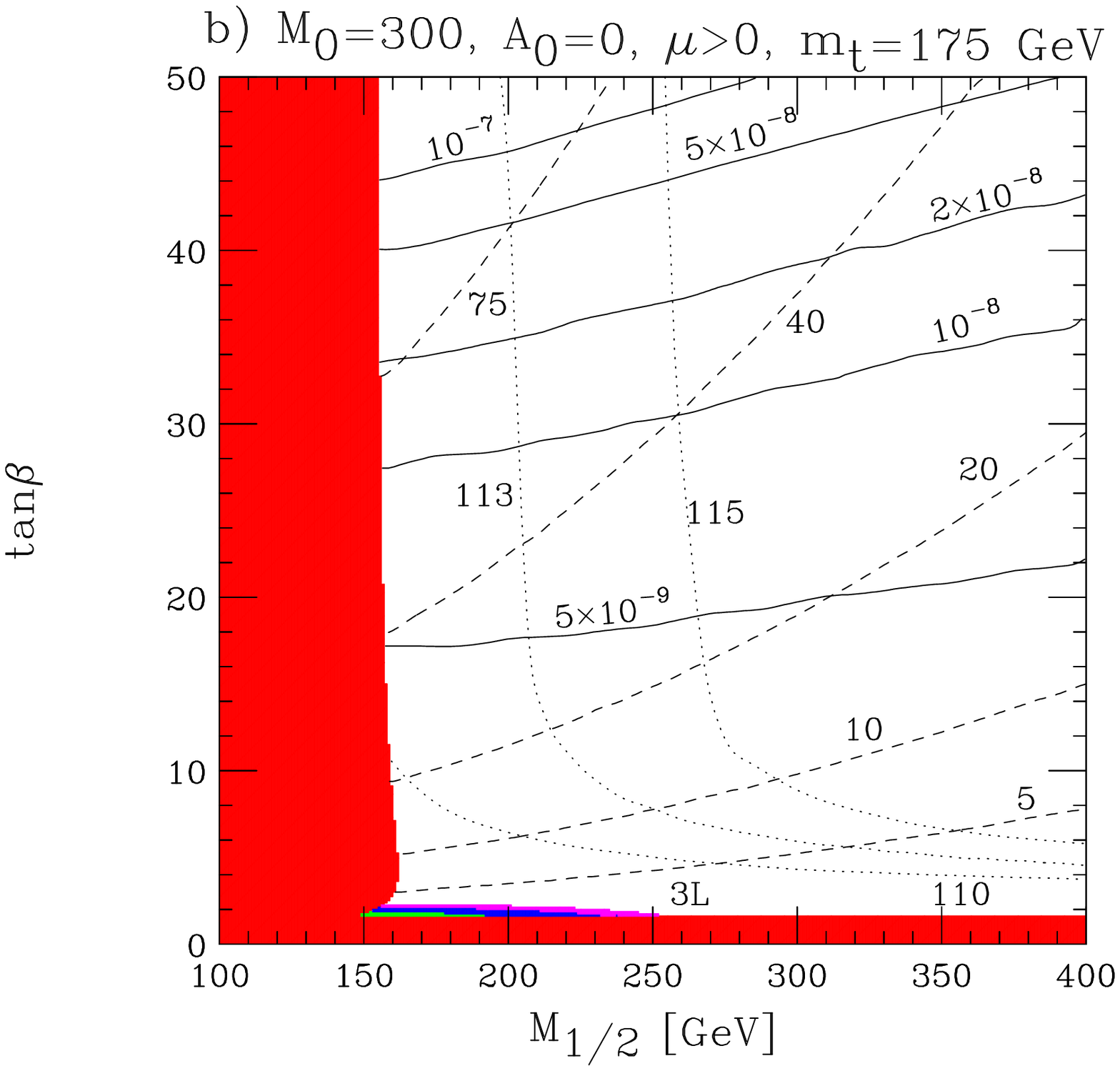,height=3in,angle=90}}}
\centerline{\hbox{\psfig{figure=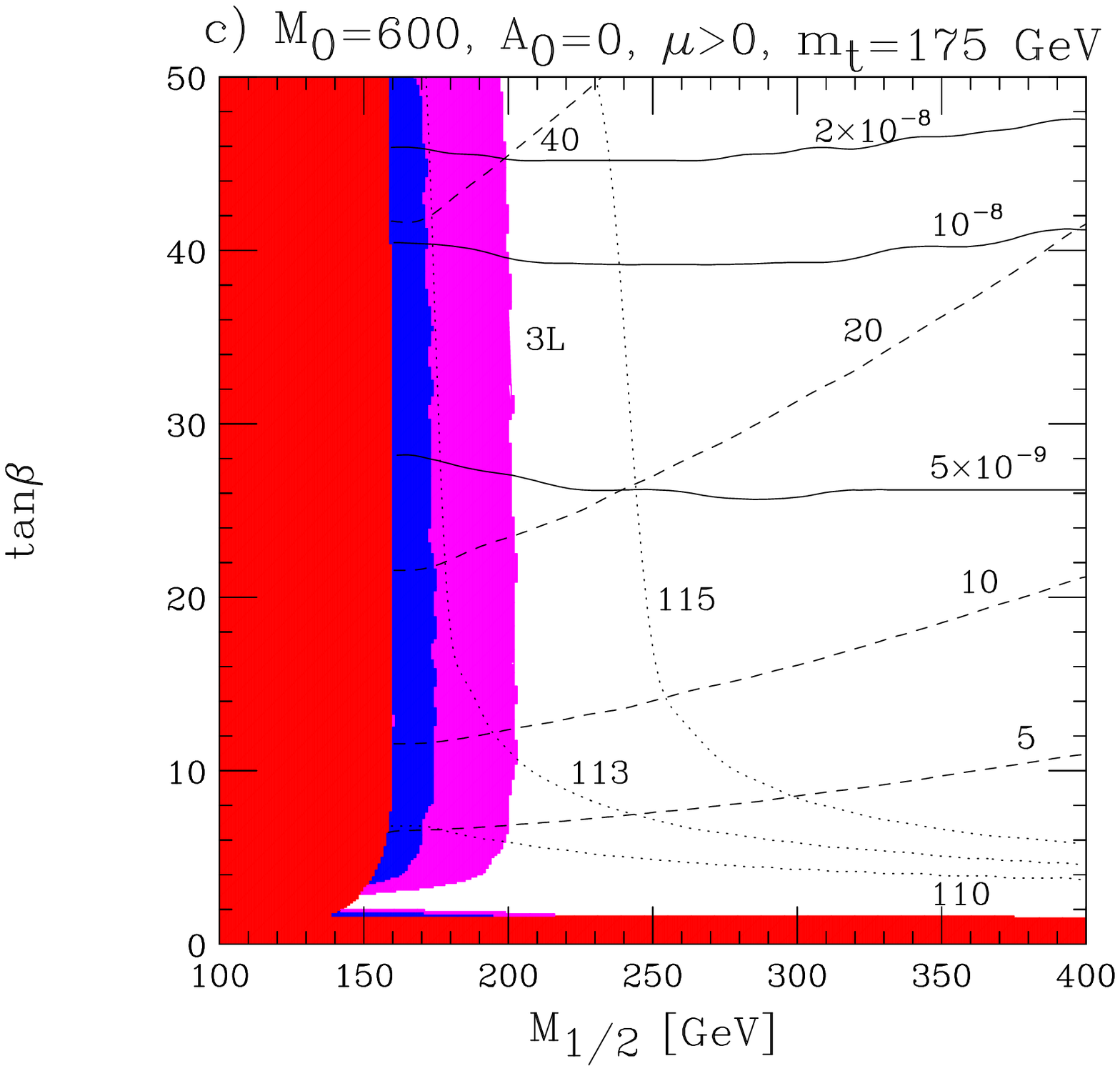,height=3in,angle=90}}}
\caption{{\it Tevatron 5$\sigma$ reach in the trilepton (3L) channel
in the $M_{1/2}-\tan\beta$ 
plane for the mSUGRA scenario with  a) $M_0=150$ GeV, b) $M_0=300$ GeV
and  c) $M_0=600$ GeV
 and $A_0=0$, $\mu>0$ and $m_t=175$ GeV. The shaded
areas indicate integrated luminosities of 30$fb^{-1}$ (magenta), 
10$fb^{-1}$ (blue) and 2$fb^{-1}$ (green) and from 
top to bottom, respectively. Dashed contours represent the SUSY 
contribution to the muon anomalous magnetic moment (in units
of $10^{-10}$) and  the dotted contours the Higgs boson mass. The
solid contours indicate the prediction
for  $\cbs$.  }}
\end{figure}

Our next step is to present the simulation of the 3L events in the
$M_{1/2}-\tan\beta$ plane for different values of $M_0$. Our results
are depicted in Fig.2. For $M_0=150$ GeV, the 5$\sigma$ reach of the
Tevatron with ${\cal L}=30 fb^{-1}$ is bounded by $\tan\beta \lsim 30$
and $M_{1/2}\lsim 270$ GeV. In this region the Higgs mass is always
below 114 GeV.  With an integrated luminosity of ${\cal L}=2 fb^{-1}$,
the Tevatron can observe the 3L events in a region with small
$\tan\beta \le 5$ and $M_{1/2}\le 210$ GeV. In this region, the
maximum Higgs mass is 109 GeV.  

As $\tan\beta$ increases, the stau mass decreases and the gaugino
decays are mediated mostly by staus rather than sleptons.  Gradually
the 3L event rate is decreasing as we increase $\tan\beta$.  When the
3L events are no longer observable, the branching ratio $\cbs$ becomes
significant and the decay $B_s\ra\mu^+\mu^-$ has an observable rate of
up to 30 events with ${\cal L}=2 fb^{-1}$.  Moving to higher values of
$M_0=300$ GeV (see Fig.2), the 3L events dissappear for the reason we
have already explained when discussing Fig.1. However, a large portion
of the parameter space with $\tan\beta \gsim 30$ is still accessible
to \bs ~ searches. The parameter space with $\tan\beta$ values below
30 has to be covered by other searches, most probably Higgs searches
(squarks and gluinos are too heavy to be directly discovered at
Tevatron in the mSUGRA scenario~\cite{mSUGRATEVREP}.)  The two cases
$M_0=150\,$GeV and $M_0=300\,$GeV have to be considered respectively
as the most ``promising'' and the ``nightmare'' benchmark scenarios
for the Tevatron Run II. 
In order to complete the discussion on $M_{1/2}-\tan\beta$ plane
we show in Fig. 2c the corresponding plot for $M_0=600$ GeV.
This region is intermediate in search sensitivity and is bounded by
$M_1/2 \lsim 210$ GeV. It is largely independent of $\tan\beta$.
A considerable amount of the remaining parameter space is covered
by the search from \bs. We observe a considerable
overlap between the two search regions for $\tan\beta \gsim 40$.
 For a complete analysis of the \bs ~mode and
other observables on $M_{1/2}-\tan\beta$ plane the reader is referred
to Refs.~\cite{sakis,Tanaka}.

\begin{figure}[t]
\centerline{\hbox{\psfig{figure=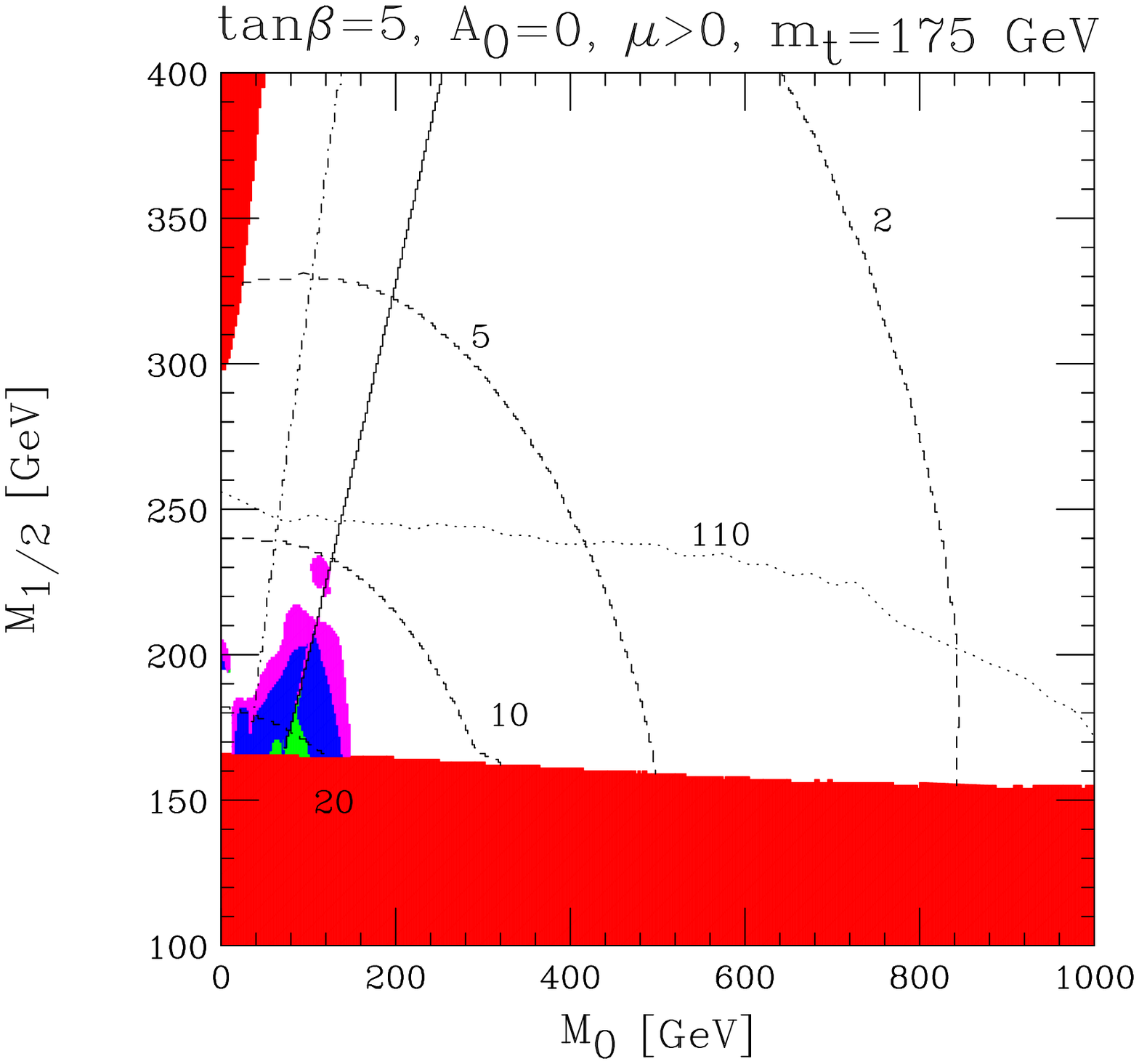,height=3in,angle=90}}
{\psfig{figure=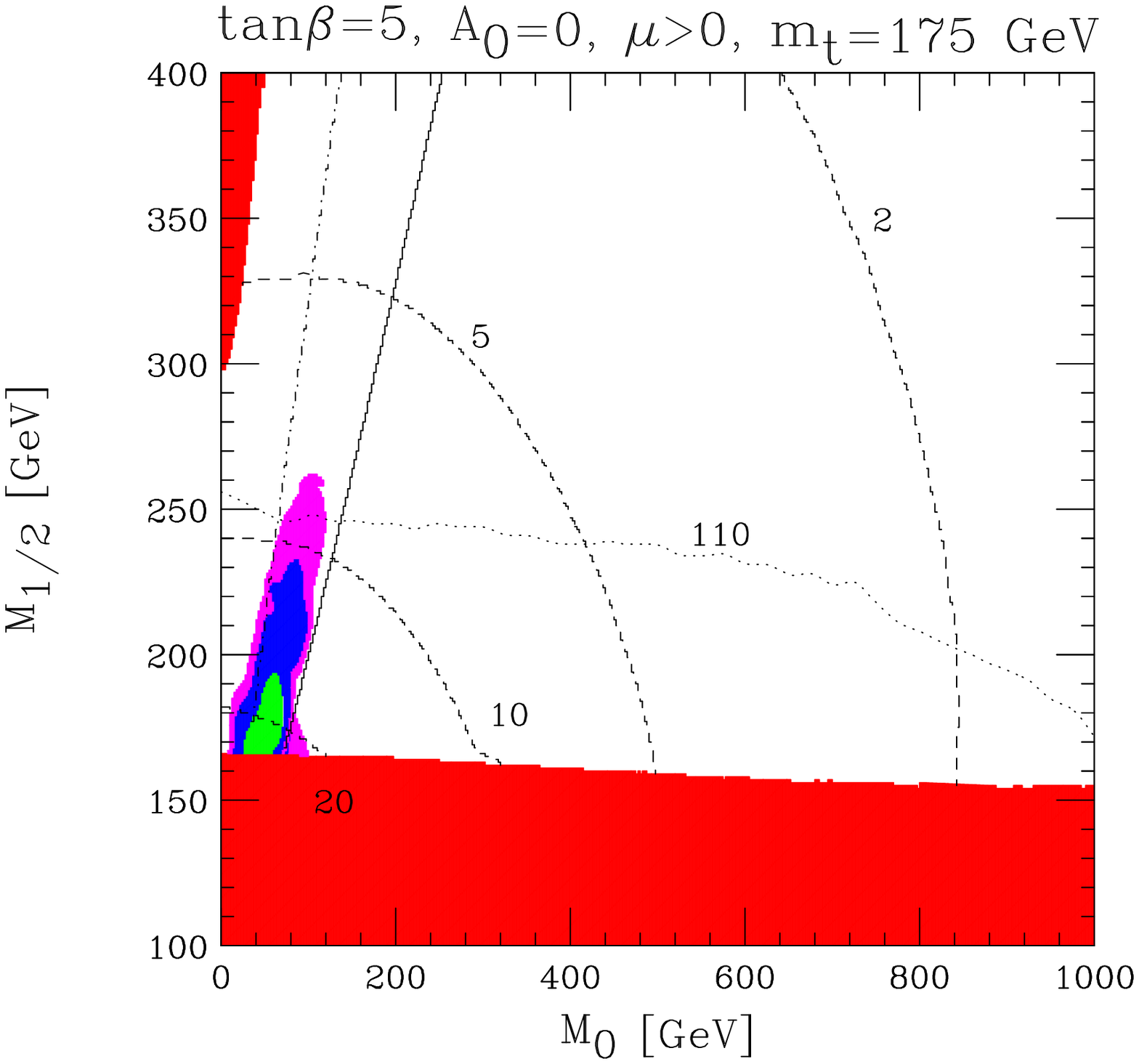,height=3in,angle=90}}}
\caption{{\it Tevatron reach in the like sign dilepton channel (2L) (left)
and and dilepton plus tau jet channel (2L1T) (right) for low $\tan\beta=5$
and all the other parameters as in Fig.1. We do not observe these two
modes at large $\tan\beta=50$. Contours and shaded areas are 
as in Fig.1.}}
\end{figure}

\subsection{Other than 3L Channels}
Apart from the trilepton (3L) analysis, we also study the modes with
like sign dileptons (2L), dilepton plus tau-jet (2L1T), lepton plus
two tau-jets (1L2T), three tau-jets (3T), like sign lepton plus
tau-jet (1L1T), and two like sign tau-jets (2T) in the final state.
The idea is to identify alternative signatures in regions of the
parameter space where the 3L signal is suppressed. For all these
channels only the 2L and 2L1T survive our cuts and only in the small
$\tan\beta \simeq 5$ region.  The results are shown in Fig.3.
Depending on the luminosity, we observe a 5$\sigma$ reach at the
Tevatron, in the mSUGRA regions:

\begin{eqnarray}
(2L, \tan\beta=5) &:&\Biggl \{\begin{array}{c} 
M_0 \lsim 120~{\rm GeV} \;\;\;,\;\;\; M_{1/2} \lsim 175~{\rm GeV} 
\;\;\;,\;\;\; {\cal L}=2 fb^{-1} \;, \\
M_0 \lsim 150~{\rm GeV} \;\;\;,\;\;\; M_{1/2} \lsim 200~{\rm GeV} 
\;\;\;,\;\;\; {\cal L}=10 fb^{-1} \;, \\
M_0 \lsim 160~{\rm GeV} \;\;\;,\;\;\; M_{1/2} \lsim 230~{\rm GeV} 
\;\;\;,\;\;\; {\cal L}=30 fb^{-1} \;, \end{array} \nonumber \\[3mm]
(2L1T, \tan\beta=5)  &:& \Biggl \{\begin{array}{c} 
40 \lsim M_0 \lsim 80~{\rm GeV} \;\;\;,\;\;\; M_{1/2} \lsim 180~{\rm GeV} 
\;\;\;,\;\;\; {\cal L}=2 fb^{-1} \;, \\
50 \lsim M_0 \lsim 100~{\rm GeV} \;\;\;,\;\;\; M_{1/2} \lsim 260~{\rm GeV} 
\;\;\;,\;\;\; {\cal L}=10 fb^{-1} \;, \\
M_0 \lsim 120 {\rm GeV} \;\;\;,\;\;\; M_{1/2} \lsim 230~{\rm GeV} 
\;\;\;,\;\;\; {\cal L}=30 fb^{-1} \;, \end{array} \nonumber
\end{eqnarray}
Comparing with the Region~I of the 3L signal, we see that an
observation of 2L or/and 2L1T will pin down the parameters
$M_0-M_{1/2}$. In the $M_{1/2}-\tan\beta$ plane and for the optimistic
benchmark $M_0=150\,$GeV, 2L events are only observable with ${\cal
  L}=10 (30) fb^{-1}$ for $M_{1/2} \lsim 190, \tan\beta \lsim 3$
($M_{1/2} \lsim 220, \tan\beta \lsim 6$).

  The search reach we obtain for signatures involving more than one hadronic
  tau are worse than those obtained previously \cite{Guchait:2001gk}.
  The main reason for this is that we started from the leading-order
  Drell-Yan production processes with additional jets generated by
  the parton-shower algorithm for W and Z production, whereas \cite{Guchait:2001gk}
  started from the W/Z+jets process. 
  The study in \cite{Guchait:2001gk} 
  was also at the parton-level rather than using fast detector simulation.
  This means we have a much larger background from W/Z with fake taus.
  This agrees with the results of \cite{Lykken:1999kp} where the signals
  with at least two electrons/muons gave the largest reach. 
  It should also be noted the we took no account of the problems of
  triggering on the multi-hadronic tau final states which will make the situation
  even worse.

\section{Conclusions}
We have presented two complementary observables to study the parameter
space of minimal supergravity at Run-II of the Fermilab Tevatron: the 
production of charginos or neutralinos decaying into trilepton (or other
than trilepton) final
states and the branching ratio of the rare decay \bs. The simulation 
of the trilepton events has been improved over previous results in
various respects. In particular next-to-leading QCD corrections have
been incorporated where available and the full momentum and spin
correlations between production and decay of the supersymmetric
particles have been taken into account. We have also presented modified cuts.
With these enhancements we find
the Tevatron reach for trilepton events increased. The two studied
observables reveal very useful complementary information on the mSUGRA
parameter space, covering the regions of small and large $\tan\beta$,
respectively. 
Our results can be read directly from the Figs.1,2. 
For example, for $M_0=150\,\gev$, $A_0=0$ and $\mu>0$ we
find that a Run-IIb with 30 $fb^{-1}$ will probe values of
$M_{1/2}$ in the range 
\begin{eqnarray}
M_{1/2} & \lsim & 250\,\gev \; - \; 70 \, \gev \times \frac{\tan \beta}{23}
\no ,  
\end{eqnarray}
which decreases with $\tan \beta$. On the contrary, if the Tevatron 
can measure $\cbs$ down to $10^{-8}$ (which requires less
than 30 $fb^{-1}$ of integrated luminosity), this will probe the 
region with $\tan \beta > 32$ and the range 
\begin{eqnarray}
M_{1/2} & \lsim &  19 \, \gev \times \tan \beta \; 
    - \; 260\,\gev 
\no   
\end{eqnarray}
for the chosen values of $M_0$, $A_0$ and $\mu$. Similar 
results can be drawn for other choices of the parameter $M_0$.

This work should be considered as a first attempt of 
relating ``direct'' SUSY searches such as the 3L events
with an ``indirect'' rare B-decay, such as \bs
at the Tevatron. 


\section*{Acknowledgments}

We would like to thank M. Kr\"{a}mer and T. Plehn for providing the
code used in \cite{Beenakker:1999xh} for next-to-leading order
electroweak gaugino production. We would like to thank F. Kr\"uger, J.
Urban for illuminating discussions on the \bs calculation. We also
like to thank K. Matchev for his valuable comments on the trilepton
analysis.  A.D. would like to thank T. Kamon for discussions on the
Tevatron 3L events, P.~Slavich and S.~Heinemeyer for useful
discussions on the updates of the Higgs boson mass calculations.
Preliminary results of this analysis have been presented by U.N. and
A.D. at the Fermilab Users' meeting and SUSY02 conference,
respectively.  A.D. would like to acknowledge financial support from
the Network RTN European Program HPRN-CT-2000-00148 ``Physics Across
the Present Energy Frontier: Probing the Origin of Mass''. P.R.
would like to thank PPARC for financial support.



\begin{thebibliography}{99}




\bibitem{mSUGRA} 
H.~P.~Nilles,
Phys.\ Lett.\ B {\bf 115} (1982) 193; Nucl.\ Phys.\ B {\bf 217} (1983)
366;
A.~H.~Chamseddine, R.~Arnowitt and P.~Nath,
Phys.\ Rev.\ Lett.\  {\bf 49} (1982) 970;
R.~Barbieri, S.~Ferrara and C.~A.~Savoy,
Phys.\ Lett.\ B {\bf 119} (1982) 343;
L.~Hall, J.~Lykken and S.~Weinberg,
Phys.\ Rev.\ D {\bf 27} (1983) 2359;
\bibitem{Soni:1983rm}
S.~K.~Soni and H.~A.~Weldon,
Phys.\ Lett.\ B {\bf 126} (1983) 215.

\bibitem{Cremmer}
E.~Cremmer, B.~Julia, J.~Scherk, P.~van Nieuwenhuizen, S.~Ferrara and L.~Girardello,
Phys.\ Lett.\ B {\bf 79}, 231 (1978);
E.~Cremmer, B.~Julia, J.~Scherk, S.~Ferrara, L.~Girardello and P.~van Nieuwenhuizen,
Nucl.\ Phys.\ B {\bf 147}, 105 (1979).

\bibitem{Girardello:1981wz}
L.~Girardello and M.~T.~Grisaru,
Nucl.\ Phys.\ B {\bf 194} (1982) 65.

\bibitem{Nilles:1983ge}
H.~P.~Nilles,
Phys.\ Rept.\  {\bf 110} (1984) 1.



\bibitem{Ibanez:fr}
L.~E.~Ibanez and G.~G.~Ross,
Phys.\ Lett.\ B {\bf 110} (1982) 215.

\bibitem{sakis}
A.~Dedes, H.~K.~Dreiner and U.~Nierste,
Phys.\ Rev.\ Lett.\  {\bf 87}, 251804 (2001)
[arXiv:hep-ph/0108037].

\bibitem{Dicus:1983cb}
D.~A.~Dicus, S.~Nandi and X.~Tata,
Phys.\ Lett.\ B {\bf 129} (1983) 451
[Erratum-ibid.\ B {\bf 145} (1984) 448].

\bibitem{Baer:1985at}
H.~Baer and X.~Tata,
Phys.\ Lett.\ B {\bf 155} (1985) 278.


\bibitem{Baer:1986dv}
H.~Baer, K.~Hagiwara and X.~Tata,
Phys.\ Rev.\ Lett.\  {\bf 57} (1986) 294.

\bibitem{Baer:1986vf}
H.~Baer, K.~Hagiwara and X.~Tata,
Phys.\ Rev.\ D {\bf 35} (1987) 1598.

\bibitem{Matchev:1999yn}
K.~T.~Matchev and D.~M.~Pierce,
Phys.\ Lett.\ B {\bf 467} (1999) 225
[arXiv:hep-ph/9907505].


\bibitem{Matchev:1999nb}
K.~T.~Matchev and D.~M.~Pierce,
Phys.\ Rev.\ D {\bf 60} (1999) 075004
[arXiv:hep-ph/9904282].



\bibitem{Buchalla}
G.~Buchalla and A.~J.~Buras,
Nucl.\ Phys.\ B {\bf 412}, 106 (1994)
[hep-ph/9308272];
M.~Misiak and J.~Urban,
Phys.\ Lett.\ B {\bf 451}, 161 (1999)
[hep-ph/9901278].

\bibitem{fbs}
C.~Bernard,
Nucl.\ Phys.\ Proc.\ Suppl.\  {\bf 94}, 159 (2001)
[hep-lat/0011064].

\bibitem{ckmfit}
\emph{CKM fitter}\ group,\\ 
{\tt http://www.slac.stanford.edu/\~{}laplace/ckmfitter/ckm\_wellcome.html}



\bibitem{L3}
W.~Adam {\it et al.}  [DELPHI Collaboration],
Z.\ Phys.\ C {\bf 72}, 207 (1996).

\bibitem{CDFbmumu}
F.~Abe {\it et al.}  [CDF Collaboration],
Phys.\ Rev.\ D {\bf 57}, 3811 (1998).


\bibitem{Nardi} 
The bound on ${\cal B} (B\to \tau^+\tau^-)$ from LEP~I 
is a byproduct of the search for $B^-\to \tau^- \bar{\nu}$ pointed out in: 
Y.~Grossman, Z.~Ligeti and E.~Nardi,
Phys.\ Rev.\ D {\bf 55}, 2768 (1997) [arXiv:hep-ph/9607473].
This bound can be improved with the existing LEP data. We would like
to thank M. Kobel and N. Wermes for discussions on this matter.



\bibitem{CLEO}
T.~Bergfeld {\it et al.}  [CLEO Collaboration],
Phys.\ Rev.\ D {\bf 62}, 091102 (2000)
[arXiv:hep-ex/0007042].




\bibitem{fnalrep}
K.~Anikeev {\it et al.},
arXiv:hep-ph/0201071.


\bibitem{Tanaka}
R.~Arnowitt, B.~Dutta, T.~Kamon and M.~Tanaka,
arXiv:hep-ph/0203069.



\bibitem{LHCb} For a review on B physics at the LHC see for
  instance,
P.~Ball {\it et al.},
hep-ph/0003238.






\bibitem{Nierste}
H.~E.~Logan and U.~Nierste,
Nucl.\ Phys.\ B {\bf 586}, 39 (2000)
[hep-ph/0004139].

\bibitem{Huang}
C.~S.~Huang, W.~Liao, Q.~S.~Yan and S.~H.~Zhu,
Phys.\ Rev.\ D {\bf 63}, 114021 (2001)
[Erratum-ibid.\ D {\bf 64}, 059902 (2001)]
[arXiv:hep-ph/0006250];
P.~H.~Chankowski and L.~Slawianowska,
Phys.\ Rev.\ D {\bf 63}, 054012 (2001)
[hep-ph/0008046].

\bibitem{Babu}
K.~S.~Babu and C.~Kolda,
Phys.\ Rev.\ Lett.\  {\bf 84}, 228 (2000)
[hep-ph/9909476]; 
G.~Isidori and A.~Retico,
JHEP {\bf 0111} (2001) 001
[arXiv:hep-ph/0110121].


\bibitem{Urban}
C.~Bobeth, T.~Ewerth, F.~Kr\"uger and J.~Urban,
Phys.\ Rev.\ D {\bf 64}, 074014 (2001)
[arXiv:hep-ph/0104284].


\bibitem{Maxim}
C.~Hamzaoui, M.~Pospelov and M.~Toharia,
Phys.\ Rev.\ D {\bf 59}, 095005 (1999)
[hep-ph/9807350].


\bibitem{Olive}
D.~A.~Demir, K.~A.~Olive and M.~B.~Voloshin,
arXiv:hep-ph/0204119.





\bibitem{BNL}
H.~N.~Brown {\it et al.}  [Muon g-2 Collaboration],
Phys.\ Rev.\ Lett.\  {\bf 86}, 2227 (2001)
[hep-ex/0102017];


\bibitem{Narison}
M.~Knecht and A.~Nyffeler,
arXiv:hep-ph/0111058.
M.~Knecht, A.~Nyffeler, M.~Perrottet and E.~De Rafael,
Phys.\ Rev.\ Lett.\  {\bf 88}, 071802 (2002)
[arXiv:hep-ph/0111059].
The bound quoted in the text is taken from
S.~Narison,
Phys.\ Lett.\ B {\bf 513}, 53 (2001)
[Erratum-ibid.\ B {\bf 526}, 414 (2002)]
[arXiv:hep-ph/0103199].




\bibitem{Baek}
S.~w.~Baek, P.~Ko and W.~Y.~Song,
arXiv:hep-ph/0205259.

\bibitem{cgnw}
L.~J.~Hall, R.~Rattazzi and U.~Sarid,
Phys.\ Rev.\ D {\bf 50} (1994) 7048
[arXiv:hep-ph/9306309];
M.~Carena, D.~Garcia, U.~Nierste and C.~E.~Wagner,
Nucl.\ Phys.\ B {\bf 577} (2000) 88 [arXiv:hep-ph/9912516].

\bibitem{Buras}
C.~Bobeth, A.~J.~Buras, F.~Kruger and J.~Urban,
Nucl.\ Phys.\ B {\bf 630}, 87 (2002)
[arXiv:hep-ph/0112305].


\bibitem{Barger:1977ap}
V.~D.~Barger, T.~Gottschalk, D.~V.~Nanopoulos and R.~J.~Phillips,
Phys.\ Rev.\ D {\bf 16} (1977) 3177.

\bibitem{Barger:1984rp}
V.~D.~Barger and R.~J.~Phillips,
Phys.\ Rev.\ D {\bf 30} (1984) 1890.



\bibitem{Chamseddine:eg}
A.~H.~Chamseddine, P.~Nath and R.~Arnowitt,
Phys.\ Lett.\ B {\bf 129} (1983) 445
[Erratum-ibid.\ B {\bf 132} (1983) 467].



\bibitem{Nath:sw}
P.~Nath and R.~Arnowitt,
Mod.\ Phys.\ Lett.\ A {\bf 2} (1987) 331.


\bibitem{Baer:1992dc}
H.~Baer and X.~Tata,
Phys.\ Rev.\ D {\bf 47} (1993) 2739.
 
\bibitem{Baer:nr}
H.~Baer, C.~h.~Chen, F.~Paige and X.~Tata,
Phys.\ Rev.\ D {\bf 50} (1994) 4508
[arXiv:hep-ph/9404212].

\bibitem{Dreiner:ba}
H.~K.~Dreiner, M.~Guchait and D.~P.~Roy,
Phys.\ Rev.\ D {\bf 49} (1994) 3270
[arXiv:hep-ph/9310291].

\bibitem{Kamon:1994yq}
T.~Kamon, J.~L.~Lopez, P.~McIntyre and J.~T.~White,
Phys.\ Rev.\ D {\bf 50} (1994) 5676
[arXiv:hep-ph/9406248].

\bibitem{Baer:1995bu}
H.~Baer, C.~h.~Chen, C.~Kao and X.~Tata,
Phys.\ Rev.\ D {\bf 52} (1995) 1565
[arXiv:hep-ph/9504234].


\bibitem{Mrenna:1995ax}
S.~Mrenna, G.~L.~Kane, G.~D.~Kribs and J.~D.~Wells,
Phys.\ Rev.\ D {\bf 53} (1996) 1168
[arXiv:hep-ph/9505245].

\bibitem{Baer:1994nc}
H.~Baer, C.~H.~Chen, R.~Munroe, F.~E.~Paige and X.~Tata,
Phys.\ Rev.\ D {\bf 51} (1995) 1046
[arXiv:hep-ph/9408265].

\bibitem{Baer:1994dq}
H.~Baer, J.~F.~Gunion, C.~Kao and H.~Pois,
Phys.\ Rev.\ D {\bf 51} (1995) 2159
[arXiv:hep-ph/9406374].

\bibitem{Lopez:1994dm}
J.~L.~Lopez, D.~V.~Nanopoulos, X.~Wang and A.~Zichichi,
Phys.\ Rev.\ D {\bf 52} (1995) 142
[arXiv:hep-ph/9412346].


\bibitem{Baer:1997yi}
H.~Baer, C.~h.~Chen, M.~Drees, F.~Paige and X.~Tata,
Phys.\ Rev.\ Lett.\  {\bf 79} (1997) 986
[arXiv:hep-ph/9704457].

\bibitem{Baer:1998bj}
H.~Baer, C.~h.~Chen, M.~Drees, F.~Paige and X.~Tata,
Phys.\ Rev.\ D {\bf 58} (1998) 075008
[arXiv:hep-ph/9802441].


\bibitem{Barger:1998wn}
V.~D.~Barger, C.~Kao and T.~j.~Li,
Phys.\ Lett.\ B {\bf 433} (1998) 328
[arXiv:hep-ph/9804451].

\bibitem{Barger:1998hp}
V.~D.~Barger and C.~Kao,
Phys.\ Rev.\ D {\bf 60} (1999) 115015
[arXiv:hep-ph/9811489].

\bibitem{Baer:1998sz}
H.~Baer, C.~h.~Chen, M.~Drees, F.~Paige and X.~Tata,
Phys.\ Rev.\ D {\bf 59} (1999) 055014
[arXiv:hep-ph/9809223].

\bibitem{Lykken:1999kp}
J.~D.~Lykken and K.~T.~Matchev,
Phys.\ Rev.\ D {\bf 61} (2000) 015001
[arXiv:hep-ph/9903238].

\bibitem{Guchait:2002xh}
M.~Guchait and D.~P.~Roy,
Phys.\ Lett.\ B {\bf 535} (2002) 243
[arXiv:hep-ph/0205015].

\bibitem{Baer:1999bq}
H.~Baer, M.~Drees, F.~Paige, P.~Quintana and X.~Tata,
Phys.\ Rev.\ D {\bf 61} (2000) 095007
[arXiv:hep-ph/9906233].

\bibitem{Baer:1999sp}
H.~Baer, F.~E.~Paige, S.~D.~Protopopescu and X.~Tata,
arXiv:hep-ph/0001086.

\bibitem{pythia}
T.~Sjostrand,
Comput.\ Phys.\ Commun.\  {\bf 82} (1994) 74;
S.~Mrenna,
Comput.\ Phys.\ Commun.\  {\bf 101} (1997) 232
[arXiv:hep-ph/9609360].


\bibitem{Pukhov:1999gg}
A.~Pukhov {\it et al.},
arXiv:hep-ph/9908288.


\bibitem{HERWIG64}
G.~Corcella {\it et al.},
JHEP {\bf 0101} (2001) 010
[arXiv:hep-ph/0011363];
G.~Corcella {\it et al.},
arXiv:hep-ph/0201201;
S.~Moretti, K.~Odagiri, P.~Richardson, M.~H.~Seymour and B.~R.~Webber,
JHEP {\bf 0204} (2002) 028
[arXiv:hep-ph/0204123].


\bibitem{Richardson:2001df}
P.~Richardson,
JHEP {\bf 0111} (2001) 029
[arXiv:hep-ph/0110108].


\bibitem{Jadach:1993hs}
S.~Jadach, Z.~Was, R.~Decker and J.~H.~K\"uhn,
Comput.\ Phys.\ Commun.\  {\bf 76} (1993) 361.


\bibitem{pgs}
PGS detector simulation package {\bf 
http://www.physics.rutgers.edu/~jconway/soft/pgs/pgs.html.}

\bibitem{Campbell:1999ah}
J.~M.~Campbell and R.~K.~Ellis,
Phys.\ Rev.\ D {\bf 60} (1999) 113006
[arXiv:hep-ph/9905386].

\bibitem{pgstau}
Talk given by K. Matchev at the  conference \emph{Higgs and Supersymmetry},
Gainesville, USA, March 7-11,1999.

\bibitem{Corcella:1999gs}
G.~Corcella and M.~H.~Seymour,
Nucl.\ Phys.\ B {\bf 565} (2000) 227
[arXiv:hep-ph/9908388].


\bibitem{Campbell:2002tg}
J.~Campbell and R.~K.~Ellis,
arXiv:hep-ph/0202176.



\bibitem{Bonciani:1998vc}
R.~Bonciani, S.~Catani, M.~L.~Mangano and P.~Nason,
Nucl.\ Phys.\ B {\bf 529} (1998) 424
[arXiv:hep-ph/9801375].

\bibitem{Beenakker:1999xh}
W.~Beenakker, M.~Klasen, M.~Kr\"{a}mer, T.~Plehn, M.~Spira and P.~M.~Zerwas,
Phys.\ Rev.\ Lett.\  {\bf 83} (1999) 3780
[arXiv:hep-ph/9906298].

\bibitem{Guchait:2001gk}
M.~Guchait and D.~P.~Roy,
arXiv:hep-ph/0109096.





\bibitem{Martin}
S.~P.~Martin and P.~Ramond,
Phys.\ Rev.\ D {\bf 48}, 5365 (1993)
[arXiv:hep-ph/9306314].





\bibitem{FeynHiggs}
For the Higgs spectrum we use the code {\tt FeynHiggs v1.2} in
M.~Frank, S.~Heinemeyer, W.~Hollik and G.~Weiglein,
arXiv:hep-ph/0202166 and references therein.  The Higgs boson mass
contours presented in the figures is slightly different than in
\cite{sakis} due to this update.  The two loop bottom Yukawa
corrections to the Higgs mass have been recently calculated in
A.~Brignole, G.~Degrassi, P.~Slavich and F.~Zwirner,
arXiv:hep-ph/0206101 .  These corrections are negligible for our
choice of the sign of the Higgsino mixing parameter.



\bibitem{Sven}
S.~Heinemeyer and G.~Weiglein,
arXiv:hep-ph/0102117.


\bibitem{LEPHiggs} [LEP Higgs Working Group Collaboration], ``Searches
  for the neutral Higgs bosons of the MSSM: Preliminary combined
  results using LEP data collected at energies up to 209-GeV,''
  arXiv:hep-ex/0107030. The result quoted here is preliminary.


\bibitem{Moroi}
T.~Moroi,
Phys.\ Rev.\ D {\bf 53}, 6565 (1996)
[Erratum-ibid.\ D {\bf 56}, 4424 (1996)]
[arXiv:hep-ph/9512396].


\bibitem{Carena}
M.~Carena {\it et al.},
arXiv:hep-ph/0010338.


\bibitem{Drees}
A.~Djouadi, M.~Drees and J.~L.~Kneur,
JHEP {\bf 0108}, 055 (2001)
[arXiv:hep-ph/0107316].


\bibitem{mSUGRATEVREP}
S.~Abel {\it et al.}  [SUGRA Working Group Collaboration],
arXiv:hep-ph/0003154.

\end{thebibliography}
\end{document}